\documentclass[useAMS,usenatbib]{mn2e}
\usepackage{epsfig}

\newcommand{\cm}{cm$^{-1}$}
\newcommand{\ai}{\textit{ab initio}}
\newcommand{\jcp}{J. Chem. Phys.}
\newcommand{\aap}{A\&A}
\newcommand{\apj}{ApJ.}
\newcommand{\apjl}{ApJL}

\newcommand{\mnras}{MNRAS.}

\title[Rotational excitation of HCN and HNC by He]{The rotational excitation of HCN and HNC by He: New insights on
the HCN/HNC abundance ratio in molecular clouds}
\author[Sarrasin et al.]{E. Sarrasin$^{1}$, D. Ben Abdallah$^{2}$, M. Wernli$^{3}$, A. Faure$^{3}$, J. Cernicharo$^{4}$ and F. Lique$^{1,5}$\thanks{E-mail:francois.lique@univ-lehavre.fr}\\
$^{1}$LOMC - FRE 3102, CNRS-Universit\'{e} du Havre, 25 rue Philippe Lebon, BP 540, 76058, Le Havre, France\\
$^{2}$Laboratoire de Spectroscopie atomique, mol\'{e}culaire et applications, Facult\'{e} des Sciences Universit\'{e} Tunis el Manar, Tunis 1060, Tunisie\\
$^{3}$Laboratoire d'Astrophysique de Grenoble, Universit\'e Joseph Fourier, CNRS UMR 5571, BP 53, 38041 Grenoble Cedex 09, France\\
$^{4}$Department of Astrophysics. CAB, INTA-CSIC. Crta Torrej\'on a Ajalvir km 4. 28850 Torrej\'on de Ardoz. Spain\\
$^{5}$LERMA and UMR 8112 of CNRS, Observatoire de Paris-Meudon, 92195 Meudon Cedex, France}
\begin{document}

\date{Accepted 2009 month XX. Received 2009 month 06; in original form 2009 month 06}

\pagerange{\pageref{firstpage}--\pageref{lastpage}} \pubyear{2009}

\maketitle

\label{firstpage}

\begin{abstract}
Modeling of molecular emission from interstellar clouds requires the
calculation of rates for excitation by collisions with the most
abundant species. The present paper focuses on the calculation of rate
coefficients for rotational excitation of the HCN and HNC molecules in
their ground vibrational state in collision with He. The calculations
are based on new two-dimensional potential energy surfaces obtained
from highly correlated \textit{ab initio} calculations. Calculations
of pure rotational (de)excitation cross sections of HCN and HNC by He
were performed using the essentially exact close-coupling
method. Cross sections for transitions among the 8 first rotational
levels of HCN and HNC were calculated for kinetic energies up to
1000~cm$^{-1}$. These cross sections were used to determine
collisional rate constants for temperatures ranging from 5~K to
100~K. A propensity for even $\Delta j$ transitions is observed in the
case of HCN--He collisions whereas a propensity for odd $\Delta j$
transitions is observed in the case of HNC--He collisions. The
consequences for astrophysical models are evaluated and it is shown
that the use of HCN rate coefficients to interpret HNC observations
can lead to significant inaccuracies in the determination of the HNC
abundance, in particular in cold dark clouds for which the new HNC rates
show that the $j=1-0$ line of this species will be more easily excited
by collisions than HCN. An important result of the new HNC-He rates 
is that the HNC/HCN abundance ratio 
derived from observations in cold clouds
has to be revised from $>$1 to $\simeq$1, in good agreement with
detailed chemical models available in the literature.
\end{abstract}


\section{Introduction}

The isomers hydrogen cyanide (HCN) and hydrogen isocyanide (HNC) are
among the most abundant organic molecules in space, from dark cold
clouds \citep[e.g.][]{Irvine1984} to circumstellar envelopes
\citep[e.g.][]{cernicharo96}, cool carbon stars \citep{harris03},
comets \citep[e.g.][]{lis97} and active galaxies \citep{perez07}. HCN
and HNC also belong to the small class of molecules detected in
high-redshift galaxies, along with CO, CN and HCO$^+$
\citep{guelin07}. In addition to thermal emission from various
rotational transitions at (sub)millimeter and far-infrared
wavelengths, a few masering lines have also been detected for both
isomers \citep[e.g.][]{lucas89,aalto09}. Thus, in contrast to purely
thermochemical considerations \citep[HNC is less stable than HCN by
  about 0.6~eV,][]{bowman93}, the abundance of HNC in space is large:
the abundance ratio [HNC]/[HCN] is observed to vary from $\sim$ 1/100,
e.g. in high mass star forming regions \citep{schilke92}, up to 4.5 in
dark cold clouds \citep{hirota98}. The large variation of this ratio
remains a puzzle in astrochemistry, although the major source of both
isomers is thought to be the dissociative recombination of HCNH$^+$
with electrons \citep[e.g.][and references therein]{amano08}. Finally,
we note that the HNC/HCN (rotational) line ratio is now employed to
distinguish photon-dominated regions (PDR) and X-ray dominated regions
(XDR): PDR sources all have ratios lower than unity while XDR have
ratios larger than 1 \citep{meijerink05}.

Due to the low density of the interstellar medium ($\sim
10^{5}$cm$^{-3}$), the rotational levels of molecules are generally
not at local thermodynamic equilibrium (LTE). Reliable determination
of molecular abundances therefore relies on accurate molecular
collisional rates. Hydrogen molecules are generally the most abundant
colliding parners in interstellar space, although collisions with H,
He and free electrons can also play important roles in energetic
regions. Rate coefficients for the rotational excitation of HCN by He
atoms (employed as substitutes for H$_2$) have been calculated by
\cite{green74}, Green (unpublished) \footnote{The earlier calculations
  of Green \& Thaddeus (1974) (restricted to the lowest 8 levels of
  HCN and temperatures of 5$-$100 K) were extended in 1993 to obtain
  rate constants among the lowest 30 rotational levels and for
  temperatures of 100$-$1200 K. These unpublished results are
  available at {\tt http://data.giss.nasa.gov/mcrates/\#hcn}} and
\cite{monteiro86} (see also references therein). All these studies
were based on the same potential energy surface (PES) obtained using
the uniform electron gas model \citep{green74}. For HNC, there is to
the best of our knowledge no collisional data except those of
\cite{faure07} for electron-impact excitation. In astrophysical
applications, HNC and HCN have been therefore assumed to present
similar collisional rates \citep[see, in particular, the discussion in
][]{guelin07}. Similar HCN and HNC rotational rates were also observed
for electron-impact excitation which is, however, dominated by the
dipole interaction \citep{faure07}. In contrast, inelastic collisions
between neutral species are dominated by short-range interactions and
larger differences between HCN and HNC are expected.

In the present work we present new rotational rate coefficients for
HCN and HNC based on highly accurate, HCN$-$He and HNC$-$He PES. The
paper is organized as follows: Section 2 describes the PES used in
this work.  Section 3 then contains a rapid description of the
scattering calculations. In Section 4 we present and discuss our
results. Finally, we analyze in section~5 the effect of these new rate
coefficients on the excitation of HCN and HNC by modeling this
excitation through a large velocity gradient code.

\section{Potential energy surface}

\subsection{HCN--He}\label{sec:pes}

Several theoretical PES for the HCN--He
system have been published during the last fifteen years:
\cite{atkins96,drucker95} and \cite{tocz01}, hereafter called
TDC01. In the present study, we used the latter surface. As a
reminder, it was computed \ai\ using the CCSD(T) method
\citep{Hampel92}, with a triple zeta (aug-cc-pVTZ) basis set and an
additional [$3s3p2d2f1g$] set of bond functions (full details on the
\ai\ procedure used can be found in TDC01). The HCN molecule was
treated as a linear rigid rotor with intramolecular distances fixed at
the equilibrium values $r_{HC}$=2.01350~bohr and $r_{CN}$=2.17923~bohr
(see discussion below in the case of HNC).The global minimum of this
PES is in the linear He--HCN configuration at $R$=7.97~bohr and has a
well depth of 29.90~cm$^{-1}$.

A semi-empirical surface \citep{harada02} is the latest PES published
for HCN-He. It used the TDC01 surface as a starting point. The choice
of the TDC01 surface seemed more adequate to us, as it is purely
\ai\ and thus consistent in the whole relevant interaction range,
while the empirical PES used the measured spectrum of the bound van
der Waals complex to refine the TDC01 surface in the region of the
attractive well, thus possibly losing precision in other regions that
could be as important for the present dynamical applications. 
A contour plot of the potential is shown in Fig. 1 of TDC01.

The PES being strongly anisotropic due to the length of the HCN rod,
we fitted the TDC01 surface with a Legendre polynomials basis
\begin{equation}
V(R,\theta)=\sum_{\lambda}V_{\lambda}(R)P_{\lambda}(\cos\theta)
\label{Vlambda}
\end{equation}
using exactly the same truncation technique as for the HC$_3$N-He
system in \cite{wernli07,wernli07b}. In this equation, $R$ is the distance from
the HCN centre of mass to the He atom, $\theta$ being the He--HCN
angle as measured from the H side of HCN. The combination of 25
polynomials ($\lambda_{max}=24$) and a cubic spline interpolation for
the radial coefficients ($V_{\lambda}(R)$) was employed to obtain a
precise fitting of the PES, the standard deviation between fitted and
\ai\ data being $\le 1$ \cm\ for all potential values lower than 300
\cm. We also tested our fit against the \ai\ values given in the lower
part of table I of TDC01, and found a difference $\le 3$ \cm\ for
all 16 points in this table.

As a first outcome of the fitting, the $V_\lambda$ with even $\lambda$
were found to be significantly lower than those of \citet{green74}
(table 3 in their article), indicating a lower (even) anisotropy of
the TDC01 surface. This is expected to have an effect in the
propensity rules, as shall be further discussed in section \ref{res}.

\subsection{HNC--He}

To our knowledge, no PES exists for the HNC-He van der Waals system.
The ground electronic state of the weakly bound HNC--He system is a
singlet with $A'$ refection symmetry. Within the ground electronic
state, the equilibrium geometry of the HNC molecule is linear. 
Therefore, HNC can be considered as a linear rigid
rotor. The HNC--He ``rigid rotor" PES is described by the two Jacobi
coordinates $R$, the distance from the centre of mass of HNC to the He
atom, and $\theta$, the angle between $\vec R$ and the HNC bond axis
$\vec r$, with $\theta=0$ corresponding to colinear HeCNH. The HN and
NC bond distances $r_{HN}$ and $r_{NC}$ were frozen at their
experimental equilibrium value $r_{HN} = 1.8813$~bohr and $r_{NC} =
2.2103$~bohr \citep{huber:79}. As demonstrated by \cite{lique07} for
the CS-He system, for non-hydride diatomic molecules two-dimensional
PES calculated for a frozen bond distance or obtained from full 3D
PES by averaging over the intermolecular ground state vibrational
wavefunction are very similar.  Consequently, in the present case we
anticipate that restricting $r_{HN}$ and $r_{NC}$ to their equilibrium
value will introduce little error into the calculated inelastic rate
coefficients.

The potential energy surface (PES) was calculated in the
supermolecular approach by means of a single- and double-excitation
coupled-cluster method \citep{Hampel92} (CCSD).  In addition, we
included perturbative contributions from connected triple excitations
computed as defined by \cite{watts:93} [CCSD(T)]. The calculations
were done with the MOLPRO 2006.1 package. For all three atoms we used
the standard correlation-consistent polarized valence-quadruple-zeta
basis sets of Dunning \citep{dunning:89} (cc-pVQZ) augmented with the
diffuse functions of $s$, $p$, $d$, $f$ and $g$ symmetries by
\cite{kendall:92} (aug-cc-pVQZ). This basis set was further augmented
by the [$3s3p2d2f1g$] bond functions optimized by \cite{cybulski:99}
and placed at mid-distance between the He atom and the HNC centre of
mass.

At all geometries the \cite{boys:70} counterpoise procedure is used to correct for basis set superposition error (BSSE).  
In this procedure the interaction energy is defined by
\begin{equation}
V(R,\theta)= E_{{\rm HNC-He}}(R,\theta)-E_{{\rm HNC}}(R,\theta)-E_{{\rm He}}(R,\theta)
\end{equation}
where the energies of the HNC and He subsystems are determined with the full (four atoms plus bond functions) basis set.

\begin{figure}
\begin{center}
\includegraphics[width=8.0cm,angle=0.]{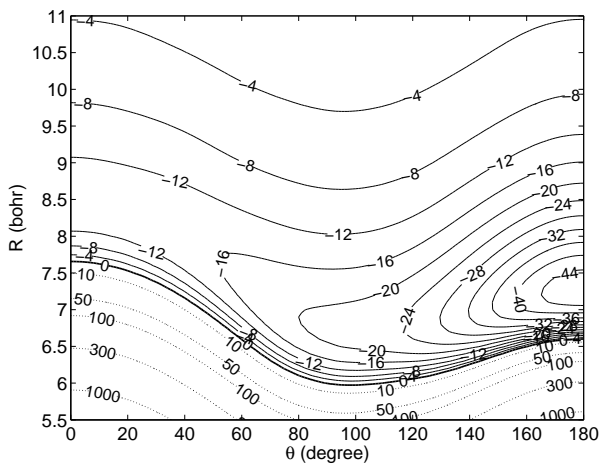}
\caption{Contour plot of the 2D HNC--He PES as a function of $R$ and $\theta$. The energies are in cm$^{-1}$ with the zero of energy taken to be HNC+He at infinite separation.}
\label{fig1}
\end{center}
\end{figure}

Interaction energies at a total of 446 geometries were computed. The values of the radial scattering coordinate $R$ ranged 
from 4 bohr to 25 bohr. The angular grid was uniform with a 15 degree spacing from 0 to 180$^\circ$. 
A contour plot of the potential is shown in Fig. \ref{fig1}. For this weakly-bound system the global minimum in the 
interaction energy was found to be --46.83~cm$^{-1}$ ($R=7.30$~bohr, $\theta=180^\circ$) corresponding to collinear He--HNC. 
We note the strong anisotropy of the PES.

The calculated interaction energies were fitted by means of the
procedure described by \cite{werner:89} for the CN--He system. The
largest deviations between the fit and the \textit{ab initio} points
occur primarily in the repulsive region of the PES. Over the entire
grid, the mean relative difference between the analytic fit and the
\textit{ab initio} calculations is 0.5\%. The dependence of the
potential energy surface on the He--HNC angle was fitted by the usual
Legendre expansion, Eq.~(\ref{Vlambda}). From an \textit{ab initio}
grid containing 13 values of $\theta$, we were able to include terms
up to $\lambda_{max}$ = 12.

\begin{figure}
\begin{center}
\includegraphics[width=5.5cm,angle=270.]{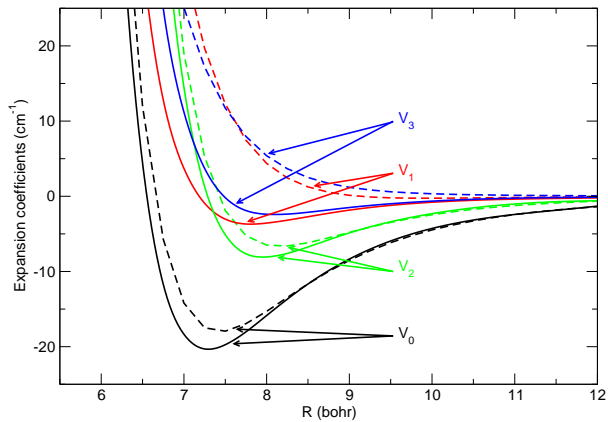}
\caption{Plot of first four radial Legendre expansion coefficients
  ($\lambda=0 ... 3$) as a function of $R$. Solid lines denote HCN--He
  while dashed lines denote HNC--He.}
\label{fig2}
\end{center}
\end{figure}

The dependence on $R$ of the dominant coefficients for the HCN--He and HNC--He PESs is shown in Fig.~\ref{fig2}. 
We observe that, for the HCN--He system, the largest (in magnitude) of
the anisotropic terms ($\lambda > 0$) corresponds to $\lambda=2$.
This implies that, to a first approximation, the PES is symmetric with
respect to $\theta\rightarrow \pi -\theta$. At the opposite, for the
HNC--He system, the largest of the anisotropic terms corresponds to
$\lambda=1, 3$ in the relevant range of $R$, that is between 7 and
8~bohrs. This reflects a large (odd) anisotropy for the
PES. Consequences for the rotational excitation will be discussed in
the next paragraph.

\section{Dynamical calculations}

The main focus of this paper is the use of the two fitted HCN--He and
HNC--He PES to determine rotational excitation and de-excitation cross
sections of HCN and HNC molecules by He atoms. In the following, the
rotational quantum number of HCN and HNC will be denoted $j$. Despite
the presence of a hyperfine structure in both molecule
\citep{garvey74}, we consider only the rotational structure of HCN and
HNC.

To determine the inelastic cross sections we used the exact close
coupling (CC) approach of \cite{arthurs:60}. The integral cross
sections are obtained by summing the partial cross sections over a
sufficiently large number of values of the total angular momentum $J$
until convergence is reached. The standard time-independent coupled
scattering equations were solved using the MOLSCAT code
\citep{molscat:94}.  Calculations were carried out at values of the
total energy ranging from 3.2 to 1000 cm$^{-1}$. The integration
parameters were chosen to ensure convergence of the cross sections. We
extended the rotational basis sufficiently to ensure convergence of
the inelastic cross sections.  At the largest total energy considered
(1000~cm$^{-1}$) the rotational basis was extended to $j=14$ and 16,
respectively, for the HCN--He and HNC--He calculations.  The maximum
value of the total angular momentum $J$ used in the calculations was
set large enough that the inelastic cross sections were converged to
within 0.005 \AA$^{2}$.

From the rotationally inelastic cross sections $\sigma_{j \to j'}
(E_{c})$, one can obtain the corresponding thermal rate coefficients
at temperature $T_K$ by an average over the collision energy ($E_c$):
\begin{eqnarray}
\label{thermal_average}
k_{j \to j'}(T_K) & = & \left(\frac{8}{\pi\mu k_B^3 T_K^3}\right)^{\frac{1}{2}}  \nonumber\\
&  & \times  \int_{0}^{\infty} \sigma_{j \to j'}(E_{c})\, E_{c}\, e^{\frac{-E_{c}}{k_BT}} dE_{c}
\end{eqnarray} 
where $\mu$ is the reduced mass and $k_B$ is Boltzmann's constant. 
To obtain precise values for the rate constants, the energy grid was chosen to be sufficiently fine to include the 
numerous scattering resonances which will be described below. 

\section{Results}\label{res} 

\subsection{Cross sections}

\begin{figure}
\begin{center}
\includegraphics[width=9cm,angle=0.]{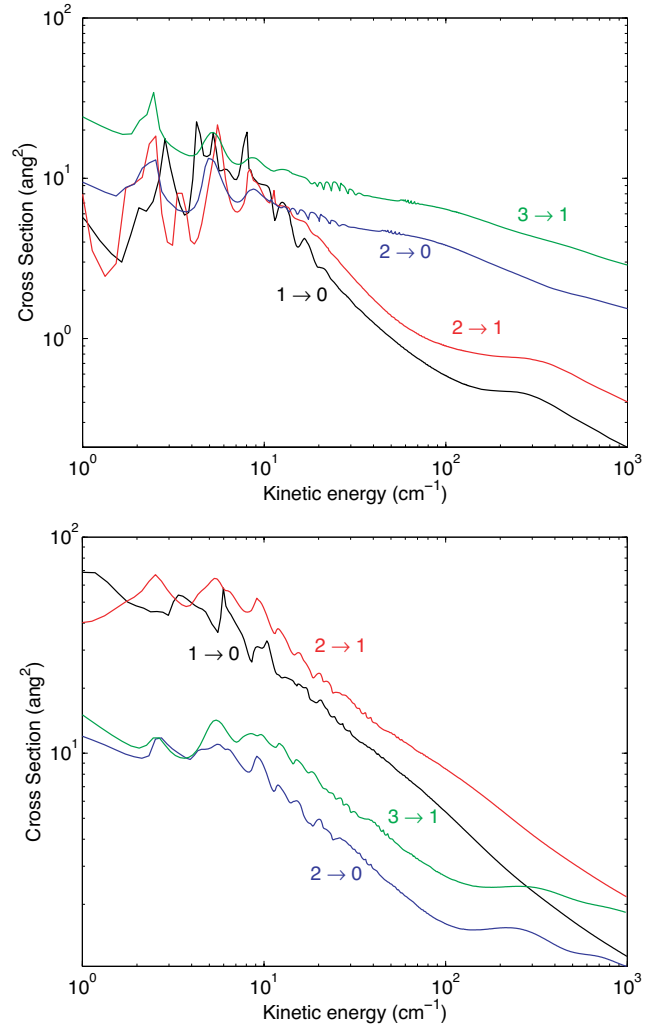}
\caption{Typical rotational de-excitation cross sections for the HCN [upper panel] and HNC [lower panel] molecules in 
collision with He as a function of the  collision energy. }
\label{fig3}
\end{center}
\end{figure}

Figure \ref{fig3} illustrates the typical energy dependence of the collisional de-excitation cross sections obtained 
from the present CC calculations for a few selected rotational levels.  

The de-excitation cross sections are almost decreasing functions of
the energy. For collision energies below 50 ~cm$^{-1}$, many
resonances are found. These are a consequence of the quasibound states
arising from tunneling through the centrifugal energy barrier (shape
resonances), or from the presence of an attractive potential well that
allows the He atom to be temporarily trapped into the well and hence
quasibound states to be formed (Feshbach resonances) before the
complex dissociates \citep{smith:79}. Because of the averaging over
collision energy, Eq.~(\ref{thermal_average}), these narrow resonances
will have little, if any, effect on the relaxation rate coefficients.

\subsection{Rate coefficients}

\begin{figure}
\begin{center}
\includegraphics[width=8cm,angle=0.]{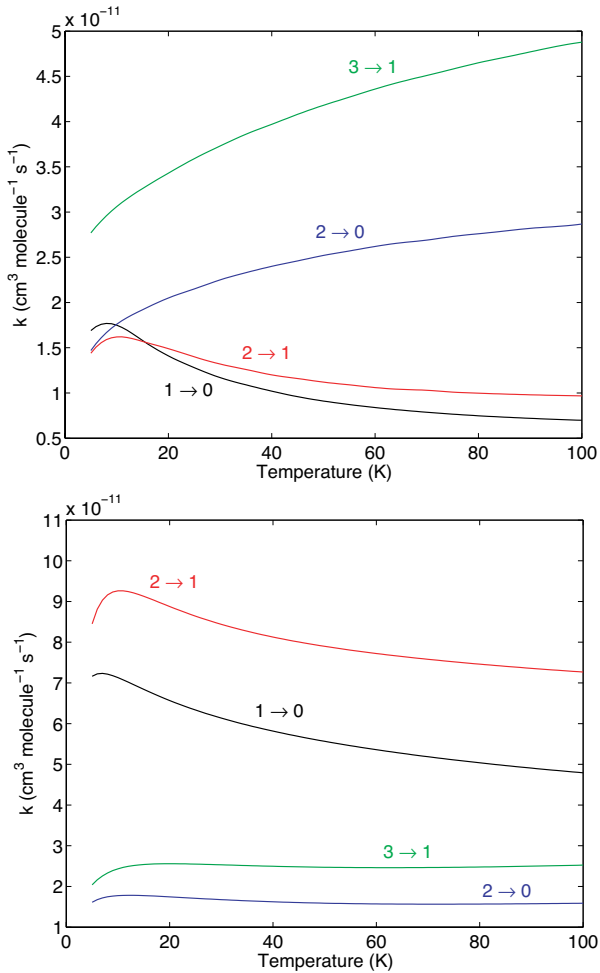}
\caption{Typical rate coefficients for the HCN [upper panel] and HNC [lower panel] molecules in collision with He as a function 
of the temperature. }
\label{fig4}
\end{center}
\end{figure}

We obtained, by energy averaging, de-excitation rate coefficients for the first 8 ($j=0-7$) rotational levels of HCN and HNC, 
from the CC cross sections.  The representative variation with temperature is illustrated in Fig \ref{fig4}. 

This complete set of (de)excitation rate coefficients is available on-line from the LAMDA 
\footnote{http://www.strw.leidenuniv.nl/~moldata/} and BASECOL\footnote{http://basecol.obspm.fr/} website. Excitation rate coefficients can be easily obtained by 
detailed balance:

\begin{equation}
k_{j \to j'}(T_K) = k_{j' \to j}(T_K) \frac{2j'+1}{2j+1} exp\left[-\left(\varepsilon_{j'}-\varepsilon_j\right)/kT_K\right]
\end{equation}
where $\varepsilon_{j}$ and $\varepsilon_{j'}$ are, respectively, the energies of the rotational levels $j$ and $j'$. 

As they are the only complete set of rotational rates available in the
literature, astronomers still generally use the HCN--He rates of
\citet{green74} in order to interpret HCN and HNC
observations. Table~1 compares on a small sample the temperature
dependence of our HCN--He rates versus those of \cite{green74}. We see
that logically, the largest differences appear at low
temperature. These differences can be attributed to the use of a new
PES for the scattering calculations. The maximum difference at 100 K -
the highest temperature for our data - is found on the smallest rate
and is of a factor of 2. On the larger rates the two sets match
nearly.

\begin{table}
  \begin{center}
    \begin{tabular}{|c|c|c|c|}
      \hline
      $j$ $j'$ & HCN  & \cite{green74} & HNC \\
      \hline
      10K && \\
      \hline
      1 0 & 0.175    & 0.110 & 0.817\\ 
      2 0 & 0.176    & 0.241 & 0.203\\
      3 0 & 0.005    & 0.012 & 0.058\\
      4 0 & 0.021    & 0.045 & 0.026\\
      \hline
      100K &&\\
      \hline
      1 0 & 0.070    & 0.074 & 0.549 \\ 
      2 0 & 0.286    & 0.285 & 0.182\\
      3 0 & 0.009    & 0.020 & 0.107\\
      4 0 & 0.064    & 0.068 & 0.031\\
      \hline
    \end{tabular}
    \caption{HCN--He and HNC--He rate coefficients at 10~K and 100~K in 10$^{-10}$~cm$^3$~molecule$^{-1}$~s$^{-1}$}
  \end{center}
  \label{table:greenrates}
\end{table}

\subsection{Comparison between HCN and HNC rate coefficients}

\begin{figure}
\begin{center}
\includegraphics[width=8cm,angle=0.]{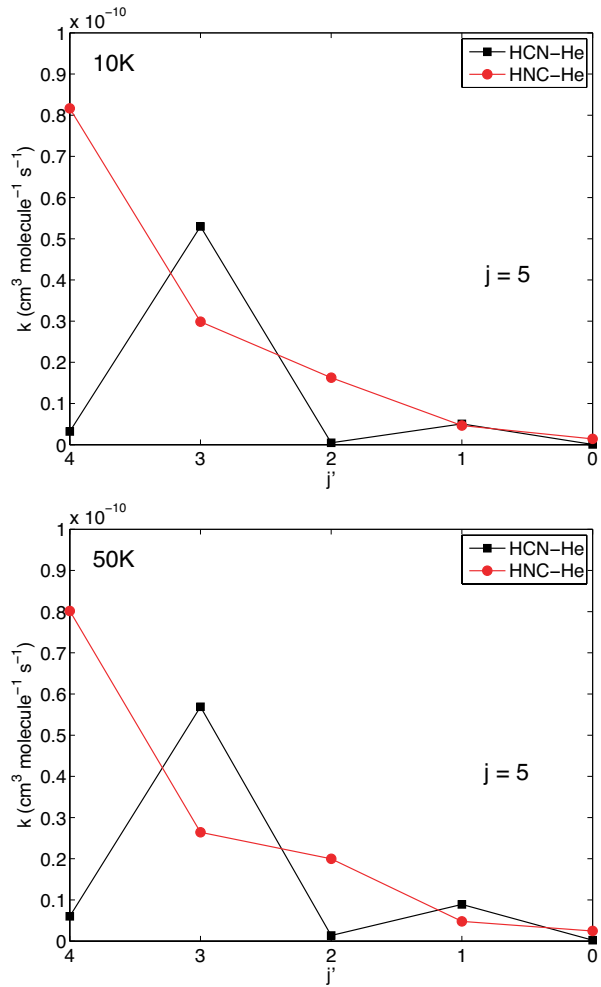}
\caption{HCN--He and HNC--He de-excitation rate coefficients from the initial level $j=5$ at 10~K [upper panel] and 50~K [lower panel].}
\label{fig5}
\end{center}
\end{figure}

Fig.~\ref{fig5} shows the HCN--He and HNC--He de-excitation rate
coefficients from the $j=5$ level at 10~K and 50~K.

One can see that significant differences exist between the HCN--He and
HNC--He rate coefficients:

(i) On the overall, the HNC rate coefficients seems to be larger than
the HCN rate coefficients. This can be explained by a larger well
depth of the HNC--He PES than the one of the HCN--He PES (46.83 versus
29.90~cm$^{-1}$).

(ii) The propensity rules seen in the two sets of rate coefficients
are different. The HCN-He rate coefficients present a strong
propensity in favour of transitions with even $\Delta j$ whereas the
HNC-He rate coefficients present a strong propensity in favor of
transitions with odd $\Delta j$. These propensity are due to the shape
of the PESs. Indeed, near-homonuclear symmetry of the potential energy
surface, such as HCN--He PES, will favor transitions with even $\Delta
j$ whereas anisotropic potential energy surface, such as the HNC--He
PES [see Fig.~(\ref{fig2})], will favor transitions with odd $\Delta
j$ \citep{mccurdy:77}.

From this comparison, one can see that the use of HCN rate
coefficients, in astrophysical applications, in order to interpret HNC
observations may be dangerous since the use of ``real HNC rate
coefficients'' will probably significantly modify the excitation of
the HNC molecule. This will be discussed in the next part.

\section{Astrophysical applications}

\subsection{HCN}
We have estimated the line intensities of HCN using the present rate
coefficients and compared the results with those obtained using the
one of \cite{green74}. Both sets of rates have been corrected for the
different reduced mass of the system HCN--H$_2$, as molecular hydrogen
is the main collision partner in molecular clouds :

 \begin{equation} \label{rdm2}
 k_{\rm HCN-H_{2}}=k_{\rm HCN-He}\left(\frac{\mu _{\rm HCN-He}}{\mu _{\rm HCN-H_{2}}}\right)^{\frac{1}{2}}
 \end{equation}
where $\mu _{\rm HCN-He}$ and $\mu _{\rm HCN-H_{2}}$ are respectivelly 
the reduced mass of the HCN-He and HCN-H$_2$ collisional systems.

\begin{figure}
\begin{center}
\includegraphics[width=8.2cm,angle=0.]{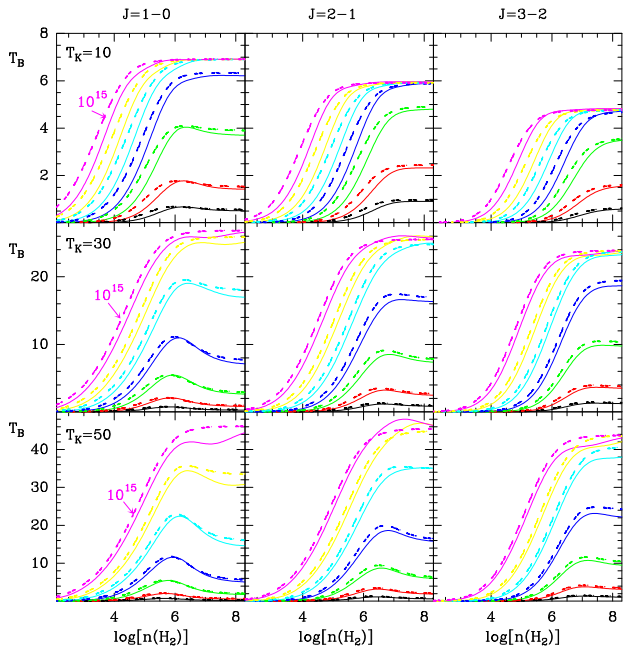}
\caption{{\bf Solid lines:} Brightness temperature for the $j=1-0$, $j=2-1$, and $j=3-2$ lines of HCN (indicated at the top
of each column) derived using our new rate coefficients for temperatures of 10, 30 and 50 K (indicated in the left panels). The column density varies
between 10$^{12}$ and 10$^{15}$~cm$^{-2}$ by steps of 10$^{1/2}$ - see text. Brightness temperatures are in K and densities in
cm$^{-3}$. {\bf Dashed lines:} The same for HNC using the HNC-He rate coefficients obtained in this work.}
\label{hcn_hnc_lvg}
\end{center}
\end{figure}

We have used a
Large Velocity Gradient (LVG) code to derive excitation temperatures,
opacities, and brightness temperatures under different physical
conditions. We have assumed a spherical cloud with an intrinsic
linewidth of 1~kms$^{-1}$. We have selected three different
temperatures: 10, 30 and 50~K, and the density has been varied from
10$^2$ to 10$^9$ cm$^{-3}$. The column densities of HCN had gone from
10$^{12}$ (optically thin case) to 10$^{15}$ cm$^{-2}$ (very optically
thick case). The results using the new rate coefficients are shown in
Fig.~(\ref{hcn_hnc_lvg}) (solid lines; dashed lines are the results
for HNC - see next section -) which can be used to estimate physical
conditions of the gas from the observed intensities of the $j=1-0$,
$2-1$, and $3-2$ lines in astronomical objects.  

In the optically thin case ($N(\rm HCN) \le 10^{13}$~cm$^{-2}$) very high densities are needed to
excite significantly the $j=1-0$ line of HCN while in the optically thick case, the HCN excitation
is dominated by radiative effects and strong emission could be obtained even in the case of densities 
$\le$~10$^4$~cm$^{-3}$. For these large opacities and narrow lines more sophisticated models 
taking into account the hyperfine structure of the molecule have to be performed
to deal in detail with the radiative transfer of HCN (see \cite{Kwan1975, Guilloteau1981, cernicharo84, 
cernicharo87,gonzalez93}).

\begin{figure}
\begin{center}
\includegraphics[width=8.2cm,angle=0.]{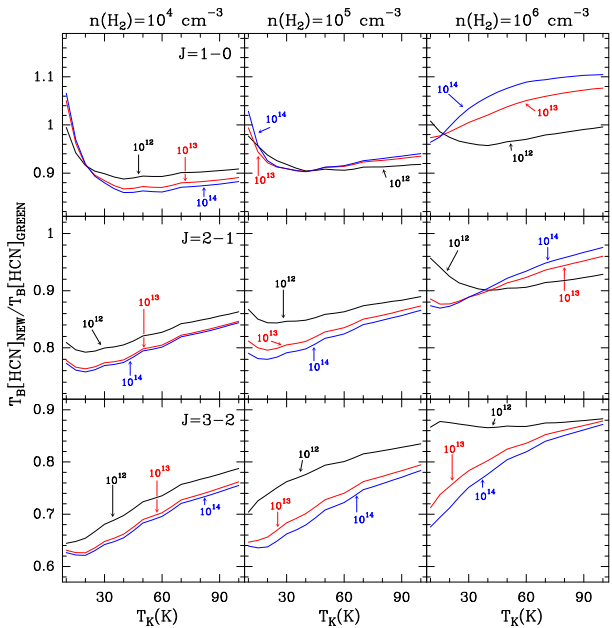}
\caption{
Ratio of the intensities of the $j=1-0$, $j=2-1$, and $j=3-2$ lines of HCN
(as indicated in the left panels) using our new collisional HCN--He rate coefficients over those 
obtained from the rates of Green \& Thaddeus (1974). 
Three densities, 10$^4$, 10$^5$, and 10$^6$~cm$^{-3}$
have been considered as indicated in the top panels. We have considered three
cases for the opacity of the $j=1-0$ line : optically thin
($N(\rm HCN)=10^{12}$~cm$^{-2}$), moderately thick ($\tau \simeq 1$,
$N(\rm HCN)=10^{13}$~cm$^{-2}$), and optically thick ($N(\rm HCN)=10^{14}$~cm$^{-2}$).
The kinetic temperature varies between 10 and 100~K (see text).}
\label{hcn_tb_ratios}
\end{center}
\end{figure}

In order to compare our results with those obtained using the
\cite{green74} rates, we show in Fig.~(\ref{hcn_tb_ratios}) the ratio
of the intensity for the $j=1-0$, $j=2-1$ and $j=3-2$ lines obtained
using the new rate coefficients and those derived using the previous
ones, $R = T_B$[New]/$T_B$[Old]. Similarly to the calculations
described above the number of rotational levels included in the
calculations is 8 ($j=0-7$). The temperature varies between 10 and
100~K, the highest temperature of the new and old HCN--He rate
coefficients. 
We have considered three different column densities for
HCN, 10$^{12}$, 10$^{13}$, and 10$^{14}$~cm$^{-2}$, which correspond
for the $j=1-0$ transition to the optically thin ($\tau <1$), moderate
optically thick ($\tau \simeq 1$), and optically thick ($\tau>$1)
cases respectively.  For the density we have selected also three
values, 10$^4$, 10$^5$ and 10$^6$~cm$^{-3}$, corresponding to low,
moderate and high collisional excitation respectively.  As one could
expect from a comparison of the rates given in Table~1, only small
differences are found in the predicted intensity ratios. For the
$j=1-0$ line we found similar intensities using both set of
collisional rates at low temperatures (a few percent difference for
T$_K$~$<$~20~K). The column density of HCN plays little effect in the
low and moderate density cases. The largest differences for the
$j=1-0$ line are found for $n$(H$_2$)=10$^6$~cm$^{-3}$.  The situation
is slightly different for the $j=2-1$ and $j=3-2$ lines. For low
temperature the intensity ratios obtained with both set of rate
coefficients show differences of 20 and 35\% respectively for low and
moderate densities (10$^4$ and 10$^5$~cm$^{-3}$).

To further explore these effects we have considered the following case:
$T_K=50$~K, $n$(H$_2$)~=~10$^5$~cm$^{-3}$ and $N(\rm
HCN)=10^{13}$~cm$^{-2}$. The old rate coefficients produce intensities
for the $j=1-0$, $j=2-1$, and $j=3-2$ lines of 3.6, 2.2 and 0.8~K
respectively. These intensities can be reproduced using the new rate
coefficients for a density of 1.3~10$^5$~cm$^{-3}$. Hence, very small
errors are introduced in the interpretation of the data using the
\cite{green74} rate coefficients for warm molecular clouds.  However,
for $T_K=10$~K the predicted intensities for these lines using the old
rates is 1.56, 0.57 and 0.09~K respectively.  With the new rates they
are 1.55, 0.45 and 0.057~K respectively. The line intensity ratios
$j=1-0$/$j=2-1$ and $j=1-0$/$j=3-2$ are different between both sets of
rates while the $j=2-1$/$j=3-2$ intensity ratio is similar. 
Hence, the
interpretation of astronomical observations of HCN using the old
HCN--He rates will be difficult in some cases as it will be hard to
fit the observed line intensity ratios. Consequently, we recommend
the use of the new HCN--He rate coefficients in interpreting 
observed line intensities.

\subsection{HNC}

Rate coefficients for HNC with Helium (or H$_2$) have not been
available until the present work. Interpretation of astronomical
observations has been done using the HCN-He rate coefficients on the
basis that the HNC-He and HCN-He PES could not be very different, that
the dipole moment of both molecules differ by less than 3\%, and that
the energy levels for both species are similar. However, the
comparison of both set of rates done in previous sections indicate
that some excitation effects could arise from the different propensity
rules for both species and from the value of the $k_{0 \rightarrow1}$
rate coefficients (see Table 1), which differs by a factor $\simeq 5$ between
both species.  We have derived brightness temperatures for the HNC
$j=1-0$, $2-1$, and $3-2$ lines using the rates calculated in this
work for the same set of models than for HCN.  The results are shown
Fig.~\ref{hcn_hnc_lvg} (dashed lines). The excitation temperature of
the $j=1-0$ line of HNC is larger than that of HCN for the three
temperatures considered in our calculations and for all column
densities (abundances). The same applies to the $j=2-1$ and $j=3-2$
transitions for low temperature. However, the differences in predicted
brightness temperatures for these two lines are less important
for kinetic temperatures above 30~K. Hence, we have two different effects for 
HNC with respect to HCN:

(i) the $j=1-0$ line will be, for identical physical conditions and
molecular abundances, considerably stronger for HNC than for HCN, i.e., it is more
easily excited which is a direct result of the larger collisional
rates for HNC. The same applies to the $j=2-1$ and $j=3-2$ lines but
the effect is less pronounced.

(ii) the intensity line ratios will have significant differences for
both molecules.

Therefore, the densities derived from astronomical observations of HNC
and HCN could be different. Moreover, in the optically thin case the
lines from these species could trace different areas of the cloud.
Our calculations indicate that the $j=1-0$ line of HNC could show weak
maser effects for moderate column densities (low line opacities) and
densities between 10$^5$ and 10$^6$~cm$^{-3}$ for temperatures above
30~K (i.e., similar to the behaviour of the HCN $j=1-0$ line).
However, no population inversion, such as the one tentatively reported
by \cite{aalto09} in the direction of Arp220, has been found for the
$j=3-2$ line.

\begin{figure}
\begin{center}
\includegraphics[width=8.2cm,angle=0.]{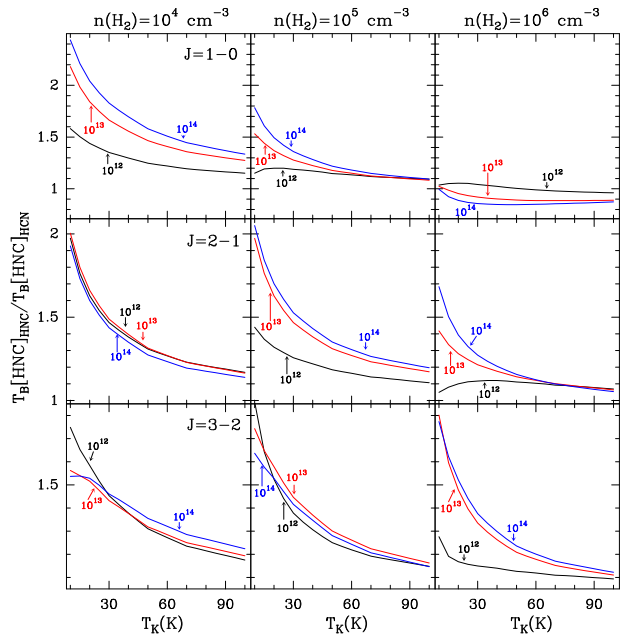}
\caption{ Ratio of the line intensities of the $j=1-0$, $j=2-1$, and
  $j=3-2$ transitions of HNC (as indicated in the left panels)
  obtained using our new collisional rates HNC--He over those obtained
  when the HCN--He rates are adopted for HNC. Three densities, 10$^4$,
  10$^5$, and 10$^6$~cm$^{-3}$ have been considered as indicated in
  the top panels. We have considered three cases for the opacity of
  the $j=1-0$ line at low kinetic temperature: optically thin ($N(\rm
  HNC)=10^{12}$~cm$^{-2}$), moderately thick ($\tau \simeq 1$, $N(\rm
  HNC)=10^{13}$~cm$^{-2}$), and optically thick ($N(\rm
  HNC)=10^{14}$~cm$^{-2}$).  The kinetic temperature varies between 10
  and 100~K (see text).}
\label{hnc_compa}
\end{center}
\end{figure}

In order to quantify the effect of the new rate coefficients on the
excitation of the rotational levels of HNC, we have compared the line
intensities obtained with the HNC-He rates and those obtained when the
HCN-He rates are adopted for HNC.  Fig.~\ref{hnc_compa} shows the
intensity ratio $R=T_B(j \rightarrow j-1)[\rm HNC-He]$/$T_B(j
\rightarrow j-1)[\rm HCN-He]$ for the $j=1-0$, $2-1$, and $3-2$
transitions. Three different column densities have been considered:
10$^{12}$, 10$^{13}$, and 10$^{14}$~cm$^{-2}$ (optically thin,
$\tau\simeq$1, and optically thick respectively). The volume density
has been selected for the case of low, medium and high collisional
excitation ($n$(H$_2$)=10$^4$, $n$(H$_2$)=10$^5$, and
$n$(H$_2$)=10$^6$~cm$^{-3}$ respectively).  We see that the intensity
ratios predicted by both sets of collisional rates approach unity for
high temperature for all column densities. However, for low and
moderate kinetic temperature there are significant differences in the
derived brightness temperatures. The $j=1-0$ transition is predicted
up to 2.5 times stronger using the HNC-He rate coefficients than that
obtained from the HCN-He ones. The effect is less prominent for the
$j=2-1$ transition and never exceeds 10\% for the $j=3-2$ one. This
behaviour has strong implications for the interpretation of the
observations of HNC in cold dark clouds as the density required to
obtain the same brightness temperature is lower for the HNC-He rates
than for the HCN-He ones. Moreover, using the latter rates it will be
difficult to fit the line intensity ratios $R_{12}=T_B(1-0)/T_B(2-1)$,
$R_{13}=T_B(1-0)/T_B(3-2)$ and $R_{23}=T_B(2-1)/T_B(3-2)$. $R_{12}$
has a value of $\ge$~7 for all column densities, n(H$_2$)$\le$10$^4$
cm$^{-2}$, and T$_K$=10 K, using the HNC-He rates. Its value is
$\simeq$~5.5 using the HCN-He ones. $R_{23}$ reachs a value of 17 with
the HNC-He rates and of 14 with the HCN-He ones for the same range of
physical conditions. For $n$(H$_2$) larger than a few 10$^6$~cm$^{-3}$
the value of $R_{jj'}$[HNC-He]/$R_{jj'}$[HCN-He] approaches unity.
For $T_K \ge $~30~K the $R_{23}$ ratio does not show any significant
difference (below 20\%) between both set of collisional
rates. However, the $R_{12}$ and $R_{13}$ ratios still show important
discrepancies for $n$(H$_2$) $\le$~10$^5$~cm$^{-3}$.

As an example to quantify how important are these errors one could 
consider a typical
dark cloud with n(H$_2$)=10$^4$~cm$^{-3}$, $T_K$=10~K and
$N$(HNC)=10$^{13}$~cm$^{-2}$. To fit the predicted $j=1-0$ line
intensity from the HNC-He rates using the HCN-He ones
we will need to increase the column density of HNC by a factor
2.4.
Consequently, it seems obvious that the HNC-He collisional rates have to be used over
the HCN-He ones at least for low and moderate temperatures. Moreover,
for low kinetic temperature the abundances of HNC reported in the
litterature have to be revised as the new rates indicated that the
HNC abundance has probably been overestimated by a factor 2-3 depending
on the physical conditions of the cloud. It is particularly interesting
to note that the $j=1-0$ line of HNC, being more easily excited than that of HCN,
could be formed in zones of the cloud different from those of HCN.

The large different collisional excitation between the $j=1-0$ line of HNC 
and HCN has important consequences for the chemical modeling
of dark clouds. \cite{Irvine1984} pointed out the need for a HNC/HCN
abundance ratio $>$ 1 in these objects. Many other authors
have arrived to the same conclusion. However, from a detailed 
modeling \cite{Herbst2000} concluded that, from a chemical point of
view, it was not possible to reproduce the derived abundances
in dark clouds and that the HNC/HCN abundance ratio could not be
very different from unity at T$_K$=10K. The present results provide 
an explanation to this observational problem and allow us to
conclude that in cold dark clouds the HNC/HCN abundance ratio
is $\simeq$1.

For $T_K\ge100$~K, it appears from Fig.~\ref{hnc_compa} that the HCN-He collisional 
rates will produce reasonable results in interpreting the observed HNC 
line intensities and line intensity ratios. Hence, the relative
abundances between both species derived in the literature toward warm 
molecular clouds are not significantly affected by the use of these rates.

\section{Summary and conclusion}

We have used quantum scattering calculations to investigate rotational
energy transfer in collisions of HCN and HNC molecules with He
atoms. The calculations are based on new, highly accurate 2D potential
energy surfaces.  Rate coefficients for transitions involving the
lowest 8 levels of these molecules were determined for temperatures
ranging from 5 to 100~K. Strong propensity rule for even $\Delta j$
were found in the case of HCN--He system whereas a propensity rules
for odd $\Delta j$ were found in the case of HNC--He system.

The impact of these new rate coefficients on astrophysical modeling
has been evaluated. The new HCN--He collisional rates have little
impact on the interpretation of observations for warm and hot
molecular clouds as the use of the previous rates \citep{green74}
produce line intensities that differ from those obtained from our
rates by an amount similar to the calibration accuracy of radio
observations in the millimeter and submillimeter domains. However, for
low kinetic temperature the previous collisional rates could not
reproduce the observed line intensity ratios $j= 1-0$/$j=2-1$ and
$j=1-0$/$j=3-2$.  For T$_K$$<$100~K we recommend to use our rates. The
situation is different for HNC as no previous collisional rates are
available and the interpretation of astronomical observations has been
based on the use of the \cite{green74} rates for HCN. We have found
significant variations in the predicted line intensities and line
intensity ratios using the HNC--He rates instead of the HCN--He ones
(our calculations for HCN). In particular, in cold dark clouds the
interpretation of the HNC line intensities will dramatically suffer if
the HCN-He rates are used rather than the correct HNC-He ones. The
main collider in molecular clouds is however molecular
hydrogen. Hence, these conclusions could be limited to the case of
HCN/HNC-para-H$_2$($j=0$) for which the collisional rates are expected to be
very similar to those of HCN/HNC-He (see below).

Our results for HNC indicate that the HNC/HCN abundance ratio derived 
from observations toward
dark clouds has to be revised. The 
best fit to the observations available in the literature
is obtained for HNC/HCN $\simeq$ 1. 
We have neglected in our calculations the hyperfine structure of HCN
and HNC. Although difficult to spectroscopically resolve for the
latter, for the former species the line intensities of the different
hyperfine components are known to show important intensity anomalies
\citep{Kwan1975, Walmsley1982, cernicharo84, gonzalez93}, which can be explained 
as the result
of their different opacities across the cloud
\citep{cernicharo84,cernicharo87,gonzalez93}. Line overlaps also play
an important role in the excitation of HCN in molecular clouds
\citep{gonzalez93}. They are responsible of the strong masers found in
other species (see, e.g., \cite{gonzalez96,gonzalez97}). Hence, a
correct treatment of the radiative transfer of HCN (and in some cases
also of HNC) in molecular clouds will require a detailed knowledge of
the collisional rates between hyperfine components. These calculations
are under development and will be published elsewhere.

Finally, as already said, the great abundance of H$_{2}$ in the
interstellar medium makes this molecule the primary collision partner
for any other species. It is generally assumed \citep{lique08b} that
rate coefficients with He can provide a good estimate of rate
coefficients for collision with para-H$_{2}$($j=0$). This
approximation postulates that collisional cross sections with He and
para-H$_{2}($j=0$)$ are equal, so that the rate coefficients differ
only by a reduced mass factor of $\approx$ 1.4 arising from the thermal
averaging (Eq.~\ref{thermal_average}). Recent results on rotational
excitation of CO \citep{wernli06}, SO
\citep{lique07b} and SiS \citep{lique08,klos08} have pointed out that
rate coefficients for collisions with para-H$_{2}$($j=0$) can be up to
a factor of 3 larger or smaller than those for collisions with He,
depending on the selected transition, but that the He rate
coefficients scaled by a factor 1.4 provide the correct order of
magnitude of the H$_{2}(j=0)$ rate coefficients\footnote{We note,
  however, that much larger differences between He and
  para-H$_2$($j=0$) have been observed for small hydride molecules
  like H$_2$O \citep{dubernet09} and NH$_3$
  \citep{maret09}. However, detailed calculations by \cite{cernicharo09} indicate
   very little differences in the water vapour predictions in protoplanetary
   disks using H$_2$O-He and H$_2$O-H$_2(j=0)$ collisional rates.}. 
Therefore, the present results should provide a
reasonable first estimate of collisional rate coefficients for
collisions of HCN and HNC with para-H$_{2}$($j=0$). On the other hand, it may
be unadvisable to use the present He rate coefficients as an estimate
of the ortho-H$_{2}$ rate coefficients, since the He and ortho-H$_{2}$
rate coefficients usually differ significantly. Specific calculations
with ortho-H$_{2}$ must be performed. Thus, it is crucial to extend
the calculations, both of the PES and of the inelastic cross sections,
to the HCN--H$_2$ and HNC--H$_2$ systems.

In any case, the full set of rate coefficients presented here will
enhance our ability to understand and interpret future HCN and HNC
observations. Collisonal rates when the hyperfine structure is
considered can be obtained from the present results and the standard
assumption that the rates will be proportional to the degeneracy of
the final hyperfine state, or employing the more physical IOS scaling
method \citep{faure07}.

\section*{Acknowledgments}
F.L. and A.F acknowledge the CNRS national program ``Physique et
Chimie du Milieu Interstellaire'' for supporting this research.
J. C. would like to thank the Spanish Ministerio de Ciencia e
Innovaci´on for funding support through grants AYA2006-14876, and the
DGU of the Madrid community government under IV-PRICIT project
S-0505/ESP-0237 (ASTROCAM).

\bibliographystyle{mn2e}


\bsp

\label{lastpage}

\end{document}